\documentclass[conference, romanappendices,10pt]{IEEEtran}
\IEEEoverridecommandlockouts

\usepackage{amsmath,amssymb,amsthm,bbm,cite,color,enumitem,graphicx,microtype,setspace,url}
\usepackage{graphicx}
\usepackage[USenglish]{babel}
\usepackage[utf8]{inputenc}
\usepackage[T1]{fontenc}
\usepackage[algo2e]{algorithm2e}
\usepackage{algorithm}
\usepackage{comment}
\usepackage{caption} 
\usepackage{soul}
\usepackage[normalem]{ulem}
\usepackage{float} 

\newcommand{\f}{\mathbf{f}}

\newcommand{\h}{\mathbf{h}}

\newcommand{\w}{\mathbf{w}}

\newcommand{\setC}{\mathcal{C}}

\newcommand{\setG}{\mathcal{G}}

\newcommand{\setN}{\mathcal{N}}

\newcommand{\Compl}{\mbox{$\mathbb{C}$}}

\newcommand{\argmax}{\operatornamewithlimits{argmax}}

\newcommand{\herm}{\mathrm{H}}

\newcommand{\refe}{\textnormal{\tiny{REF}}}
\newcommand{\comb}{\textnormal{\tiny{Comb}}}

\newcommand{\snr}{\gamma}

\newcommand{\joint}{\textnormal{\tiny{Joint}}}
\newcommand{\independent}{\textnormal{\tiny{Ind}}}

\usepackage{titlesec} 
\let\subparagraph\relax
\titlespacing{\section}{0pt}{6pt plus 2pt minus 2pt}{5pt plus 1pt minus 2pt}
\titlespacing{\subsection}{0pt}{5pt plus 2pt minus 2pt}{5pt plus 1pt minus 2pt}

\usepackage[labelsep=period,font=small]{caption}
\usepackage{tikz}
\usetikzlibrary{arrows, arrows.meta, backgrounds, calc, intersections, decorations.markings, decorations.pathreplacing, positioning, shapes}

\usepackage{pgfplots}
\usepgfplotslibrary{groupplots}
\pgfplotsset{compat=1.17}

\title{Coordinated Multi-BS SSB Beam Design for Enhanced Initial Access Coverage}
\author{
\IEEEauthorblockN{Hakime BarghiZanjani, Bikshapathi Gouda, and Antti Tölli}
\IEEEauthorblockA{Centre for Wireless Communications, University of Oulu, Finland \\
Emails: \{hakime.barghizanjani, bikshapathi.gouda, antti.tolli\}@oulu.fi \vspace{-2mm}
\thanks{This work is supported by the Academy of Finland (346208 6G~Flagship, and 357504 EETCAMD).}}\vspace{-7mm}}
\begin{document}

\maketitle

\begin{abstract}
Ensuring strong synchronization signal block (SSB) coverage is essential for reliable user equipment (UE) connection during initial access. While techniques such as power boosting and network densification are commonly used, this work explores joint transmission (JT) of SSBs as an alternative to enhance coverage. Although JT is widely applied in data transmission, its use for SSBs has not been explored due to the lack of channel state information, which prevents coherent signal alignment across base stations (BSs). To address this, we propose a repetition-based JT strategy using a small set of predefined phase configurations at the BSs. This enables the UE to coherently combine multiple SSB receptions and achieve constructive gain regardless of its location. To reduce overhead, a limited number of joint beam configurations is selected to maximize the coverage. Simulation results under the line-of-sight conditions show up to $6$~dB relative SNR gain with JT of SSBs using $4$ BSs, compared to independent SSB transmission under the same resource budget. These results highlight the potential of JT to improve the coverage of SSBs during the initial access.
\end{abstract}

\section{Introduction}
Initial access enables user equipment (UE) to discover and connect to a serving cell when entering a new area or transitioning from idle to connected mode. This involves detecting a base station (BS), acquiring system information, and initiating the connection setup process.
Extensive research has addressed initial access in 5G new radio (NR), focusing on aspects such as beam management, synchronization, random access optimization, and machine learning (ML)-based enhancements. For example, improved cell search through advanced beam sweeping and alignment was studied in~\cite{Ras21,Tom20}, while ML techniques for beam selection and access were explored in~\cite{Sin24,Dre23,Dre22}. Extensions to non-terrestrial networks were considered in~\cite{Saa19_1,Saa19_2}. While these works contribute to various stages of the initial access process, this paper focuses on improving the synchronization signal block (SSB) coverage for more reliable network detection.

The SSB is central to 5G NR initial access, allowing UEs to synchronize and receive basic system information. Unlike 4G wide-area synchronization, 5G NR uses directional transmission to combat high-frequency path loss. Beamforming enables narrow SSB beams that are swept across directions to ensure full coverage~\cite{Dah20}. However, as the number of beams increases, so does access delay, especially in higher bands requiring finer angular resolution. To address this, many works have aimed to reduce the SSB beam count while maintaining coverage. For instance, \cite{Per20} proposed an adaptive SSB allocation based on past UE detection. 
However, such methods assume independent SSB transmission by each BS. However, the potential of joint transmission (JT) for SSB transmission has not been explored.

In JT, multiple BSs coordinate to transmit the same signal simultaneously to a UE, with the objective of improving signal quality via constructive alignment. This alignment is essential for improving the signal-to-interference-plus-noise ratio (SINR) of the cell-edge UEs. JT has been widely adopted in the connected mode of 4G and 5G systems, where numerous studies, including~\cite{Dav13} have demonstrated its benefits.
However, applying JT during the initial access stage remains largely unexplored, primarily due to the absence of channel state information (CSI), which makes signal alignment across BSs infeasible. To the best of our knowledge, this work is among the first to investigate the use of JT for enhancing the SSB coverage during initial access, offering a novel approach to improving the coverage in cellular networks.

In this paper, we propose a strategy for initial access that enables constructive reception of jointly transmitted SSBs using fixed phase combinations at the BSs, without requiring the CSI. To control signaling overhead, phase alignment is restricted to a small set of complementary values, and a Hadamard-like phase selection is used to minimize the number of transmissions. The UE coherently combines all received SSBs across different phase configurations to enhance signal strength. The beam directions at the BSs are selected using a greedy approach to maximize the spatial coverage. Additionally, the number of phase combinations is reduced based on the location by selecting only the dominant contributing BSs, thereby limiting the redundancy of SSB transmissions. Simulation results for JT of SSBs with $4$ BSs show that the proposed scheme achieves up to a $6$~dB relative SNR gain over independent SSB transmission, even under equal resource budgets.
\section{Initial Access: Overview and Coverage Analysis}
In this section, we first provide a brief overview of the initial access procedure in 5G, followed by a discussion on the coverage with JT of SSBs during the initial access.

 \subsection{Overview of Initial Access in 5G}
Initial access is the process by which a UE detects and connects to a cellular network. In 5G NR, this begins with the BS periodically sweeping the SSBs across the coverage area, while the UE searches for them to detect the network. Each SSB consists of a primary and a secondary synchronization signal, which enable the UE to achieve time and frequency synchronizations and identify the cell. After detecting a valid SSB, the UE decodes the associated system information block to retrieve essential system information before initiating the random access procedure. It then transmits a preamble in the uplink to connect to the network. After a successful connection, data is exchanged between the BS and the UE. 

In 5G NR, the SSBs are transmitted directionally by each BS using beams defined by predefined codebooks. To ensure full spatial coverage, the BS performs beam sweeping, sequentially transmitting SSBs across different directions. While this directional approach improves signal detection at higher frequencies, it also increases latency, as the UE must scan multiple beams to detect a suitable SSB. Moreover, at mmWave and higher frequencies, the effectiveness of SSBs is further challenged by severe path loss, making coordinated beam design essential to maintain reliable and timely initial access. In the following, we discuss initial access coverage based on a reference SNR, considering the JT of SSBs.

\subsection{Initial Access Coverage with JT of  SSBs}
\begin{figure}[t] 
    \centering
    \includegraphics[width=1\linewidth, trim={4cm 0cm 0cm 0cm}, clip]{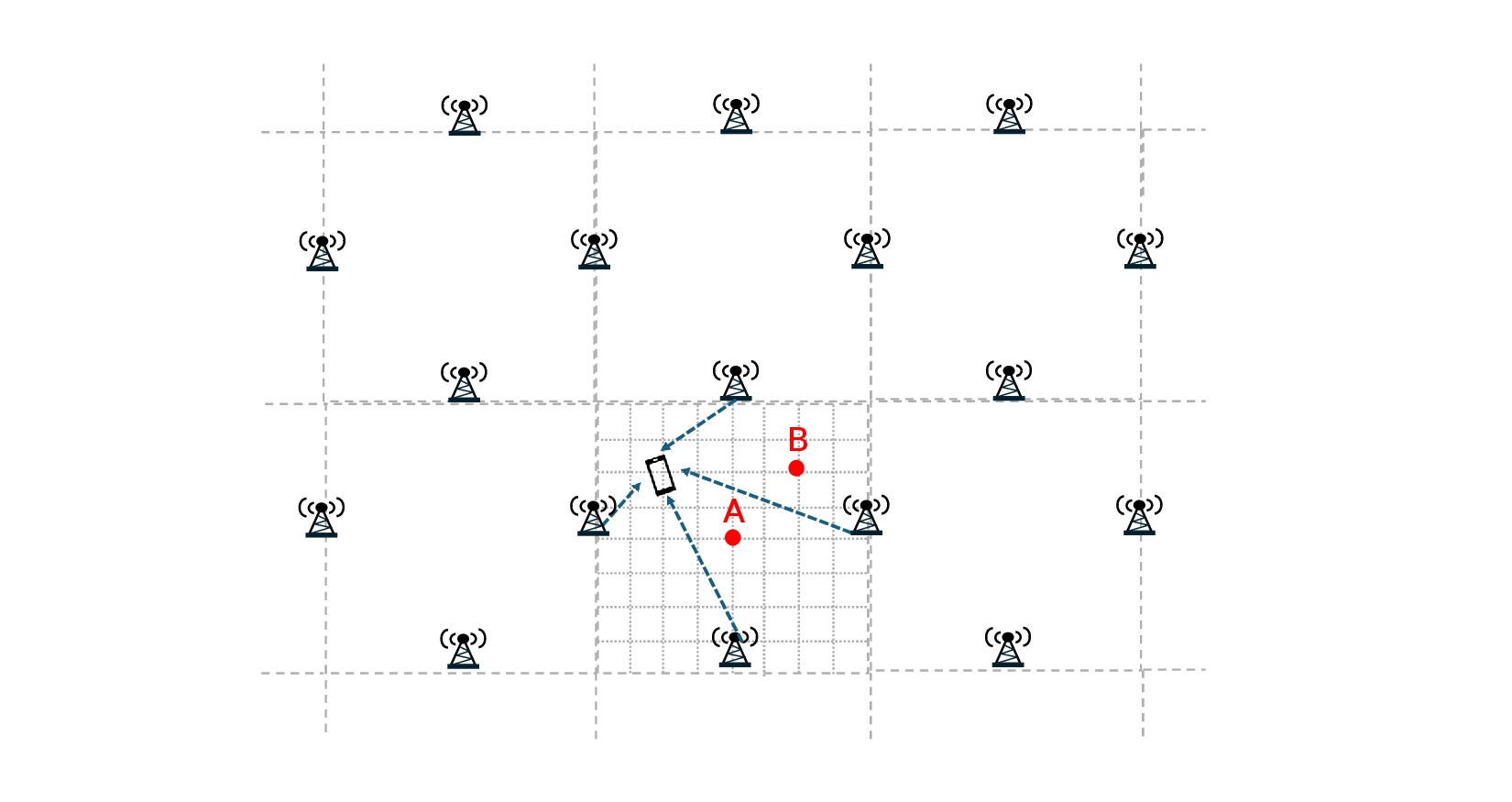}
    \caption{\small{Network topology, and coherent transmission of SSB.}} 
    \label{fig:NetworkTopology} 
    \vspace{3mm}
\end{figure}

Instead of each BS independently sweeping its own SSBs, multiple BSs can cooperatively transmit the same SSB to boost the received SNR at the UE, as illustrated in Fig.~\ref{fig:NetworkTopology}. Assuming $B$ BSs, each equipped with $N$ antennas, are coherently transmitting the SSBs, the received signal at an arbitrary UE antenna location is given as
\begin{align}\label{eq:UeRx}
    y \triangleq \sum_{b=1}^{B} \sqrt{\rho_b} \h_b^{\herm} \w_b s + n,
\end{align}
where  $\rho_b$ is the transmit power of BS~$b$, and $\h_b \in \Compl^{N \times 1}$ is the 
channel from BS~$b$ to the UE antenna, which depends on the location of the UE.\footnote{We assume single antenna UEs for simplicity, but the approach applies to multi antenna UEs, using standard combining techniques at the UE.} $\w_b \in \Compl^{N \times 1}$ is the transmit beamforming vector used at BS~$b$, $s$ is a transmitted symbol during the SSB, and $n \sim \setC\setN(0, N_0)$ is the AWGN at the UE.

Ideally, the beamforming vector $\w_b$ would be derived from the channel between the BS and a known UE. However, SSBs are intended for connecting unknown UEs located at arbitrary positions within the network area. As a result, the beamforming vectors $\w_b$ used for SSB transmission are selected from a predefined codebook. For instance, 5G employs a DFT-based codebook, as discussed earlier. In contrast to independent SSB transmission, JT from multiple BSs ideally requires phase alignment to coherently combine the signals at the UE placed at a random location and enhance the received SNR. To achieve this, each BS employs a DFT-based precoder for SSB transmission combined with appropriate (fixed) phase adjustments. 
Accordingly, the received signal at the UE expressed in~\eqref{eq:UeRx} can be rewritten as
\begin{align} \label{eq:received_signal}
y = \sum_{b=1}^{B} \sqrt{\rho_k}\h_b^{\herm} \f_b e^{i\theta_b} s + n,
\end{align}
where $\f_b \in \Compl^{N \times 1}$ is the beamforming vector selected from a DFT codebook, and $\theta_b$ is a given phase adjustment applied at BS~$b$. Lastly, the signal-to-noise ratio (SNR) at the UE for the fixed beamformers and phases at the BSs is given as:
\begin{align}\label{eq:snr_joint}
    \snr^{\joint} = \frac{1}{N_0} \left\| \sum_{b=1}^{B} \sqrt{\rho_b}\h_b^{\herm} \f_b e^{i\theta_b} \right\|^2.
\end{align}

We assume that a UE can detect the SSBs if its received SNR exceeds a predefined threshold, denoted by $\snr^{{\refe}}$, for any beamformer $\f_b$ and phase $\theta_b$ used at the BS~$b$. Consequently, using the SNR expression in~\eqref{eq:snr_joint} and assuming that the UE is uniformly distributed across the network area, the coverage probability for initial access can be defined as:
\begin{align}\label{eq:pr_cov}
    \mathbb{P}(\snr^{\joint}_{\max}(\boldsymbol{a}) \ge \snr^{\refe}) 
    \triangleq \frac{1}{|\mathcal{A}|} 
    \int_{\boldsymbol{a} \in \mathcal{A}} 
    u(\snr^{\joint}_{\max}(\boldsymbol{a}) - \snr^{\refe}) \, d\boldsymbol{a},
\end{align}
where $\snr^{\joint}_{\max}(\boldsymbol{a}) \triangleq \max_{\{\f_b, \theta_b\}} \snr^{\joint}(\boldsymbol{a})$ is the maximum SNR achieved at location $\boldsymbol{a}$, which depends on the channel vector $\h_b$ at that location. The set $\mathcal{A}$ denotes the total network area, and $u(\cdot)$ is the unit step function defined by $u(x) = 1$ if $x \ge 0$ and $u(x) = 0$ otherwise. For brevity, we may omit the argument $\boldsymbol{a}$ in the remainder of the paper.

As seen earlier, $\snr^{\joint}$ depends on both the beamforming vectors ${\mathbf{f}_b}$ and the transmission phases ${\theta_b}$. Let $\mathcal{F}_b$ denote the DFT beam codebook for BS $b$, and let $\Theta$ be the set of allowable phase values. A joint SSB configuration is defined by selecting one beam $\mathbf{f}_b \in \mathcal{F}_b$ and one phase $\theta_b \in \Theta$ for each BS. Formally, each configuration can be represented as the tuple:
\begin{align}
\boldsymbol{\psi} \triangleq (\mathbf{f}_1, \dots, \mathbf{f}_B, \theta_1, \dots, \theta_B).
\end{align}
Assuming $\mathcal{F}_b \in \mathbb{C}^{N \times N}$ and $|\Theta| = t$, each BS has $Nt$ possible beam-phase pairs, resulting in a total of $(Nt)^B$ joint SSB configurations across $B$ BSs. This exponential growth quickly becomes impractical due to signaling overhead. To address this, we propose a resource-efficient strategy that limits the number of transmitted SSBs through structured selection of beam and phase configurations. The next section presents methods for achieving this while maintaining coverage.

\section{Joint SSB Beamforming Design under Resource Constraints} 
In practice, the coverage area is discretized into a finite set of evaluation points to enable numerical computation  of the performance metric defined in~\eqref{eq:pr_cov}. Accordingly, the area $\mathcal{A}$ is divided into grid cells, where each grid has a size much smaller than the wavelength $\lambda$. The SNR is evaluated at the center of each grid cell, denoted by $g \in \setG$, and the coverage probability can be approximated as
\begin{align}\label{eq:pr_cov_dis}
    \mathbb{P}(\snr^{\joint}_{\max} \ge \snr^{\refe}) \simeq \frac{1}{|\setG|} \sum_{g \in \setG} u(\snr^{\joint}_{\max} - \snr^{\refe}), \nonumber
\end{align}
In addition to maximizing the coverage probability, practical constraints must be taken into account. Specifically, the number of jointly transmitted SSBs, i.e., the number of beam and phase combinations, should be minimized to reduce the signaling overhead and initial access delay. Based on these considerations, the optimization problem to design the joint SSB beams can be formulated as
\begin{equation} \label{eq:Opt}
\left\{
\begin{aligned}
    \max \ & \frac{1}{|\setG|} \sum_{g \in \setG} u(\snr^{\joint}_{\max} - \snr^{\refe}), \\
    \text{s.t. } & \mathcal{C} \triangleq \left\{ \psi \mid \f_b \in \mathcal{F}_b,\ \theta_b \in \Theta \right\},
    \ |\mathcal{C}| \leq N^{{\refe}},
\end{aligned}
\right.
\end{equation}
where $N^{{\refe}}$ is a design parameter that sets an upper bound on the number of jointly transmitted SSB configurations. 
Since both the beam directions and phase shifts are selected from discrete sets, solving the problem directly is challenging. Therefore, we first address the optimization of phase shifts across the BSs, followed by the optimization of beam combinations.

 \captionsetup{skip=2mm} 
\begin{figure}[!t] 
    \centering
    \includegraphics[width=0.8\linewidth, trim={0cm 0cm 0cm 0cm}, clip]{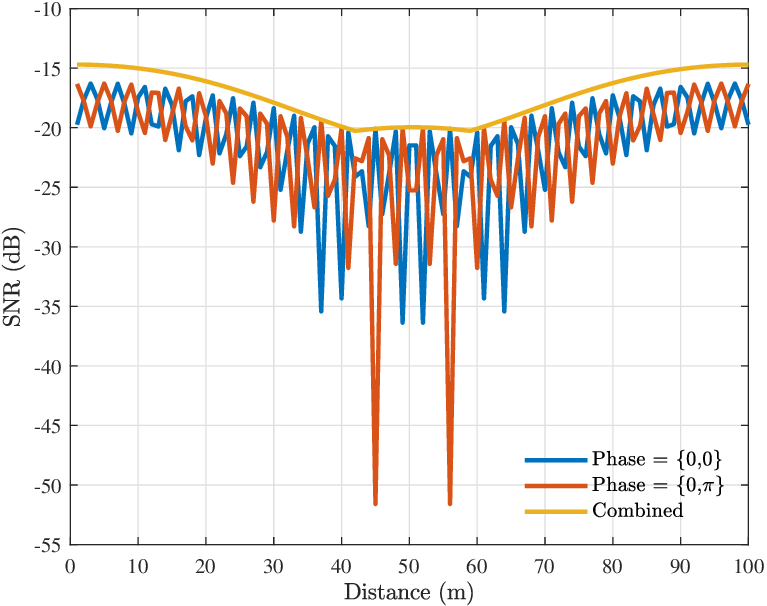}
    \caption{\small{Location-dependent constructive and destructive combining with JT of SSBs using two BSs.}}    
    \label{fig:Fading}
    \vspace{3mm}
\end{figure}
\subsection{Phase Selection}
In coordinated multi-BS transmission, achieving constructive signal addition at the UE typically requires precise phase alignment across the BSs. However, during initial access, CSI is not available, making such alignment infeasible or overly complex, especially when considering all possible channel realizations across the network. To address this, we propose a simple phase design strategy that does not rely on prior CSI. Phase values $\theta_b$ of all the BSs are defined relative to a reference BS, and without loss of generality, we set $\theta_1 = 0$, so that all other phases represent relative offsets. Consider a simple two-BS scenario with LoS channel, where the first BS (reference) has a fixed phase of $0$, and the second BS also transmits with phase $0$ during the first transmission. This setup leads to constructive signal addition at certain locations and destructive signal cancellation at others. In the next transmission, the second BS switches its phase to a complementary value, i.e., $\pi$ while keeping the beamformer unchanged, effectively flipping the phase combination, resulting in locations that were previously received constructive signal becoming destructive, and vice versa. If the UE coherently combines the two transmissions, the resulting signal benefits from constructive addition across the entire area, making the effective SNR spatially uniform and dependent only on path loss from the BSs, rather than on instantaneous phase alignment. This effect is illustrated in Fig.~\ref{fig:Fading}, where two BSs placed $100$ m apart produce alternating received signal patterns along the line connecting them. When both transmissions with phases $0$ and $\pi$ are combined, the SNR becomes spatially uniform and independent of the received signal pattern, highlighting the robustness of the proposed phase diversity approach. 
For the case of two BSs, the possible complementary phase sequences can be defined as:
\begin{align}
\Theta = \left\{ (0, 0),\ (0, \pi) \right\},
\label{eq:2Tuple}
\end{align}
and the corresponding amplitude, i.e., $e^{i\theta_b}$ values are $\left\{ (1, 1),\ (1, -1) \right\}$, which forms orthogonal sequence between the BSs. This concept can be straightforwardly extended to any number of BSs. As the number of participating BSs increases to $B$, the number of phase sets also increases to $B$, where the amplitude values form an orthogonal sequence between the BSs, like the rows of a Hadamard matrix of order $B$. The orthogonality of the Hadamard rows ensures that, when the UE coherently combines the repeated transmissions, the signals from multiple BSs constructively combine at each location, without interference from cross terms. Increasing the number of cooperating BSs leads to a higher number of repeated transmissions using the same beamformer with different phase combinations. However, the SNR contribution from distant BSs is typically negligible. Therefore, to balance the trade-off between performance gain and retransmission overhead, we limit the maximum cooperation to the four closest neighboring BSs. For this 4-BS scenario, the complementary set of phase combinations is given by:
\begin{align}
\Theta = \left\{
(0, 0, 0, 0),\ (0, \pi, \pi, 0),\ (0, \pi, 0, \pi),\ (0, 0, \pi, \pi)
\right\}.
\label{eq:4Tuple}
\end{align}
 Assuming the UE combines the received SSBs transmitted with the same beamformer but different phase settings, the resulting SNR is given as 
\begin{align}
\snr^{\joint-\comb} &= \frac{1}{N_0} \sum_{(\theta_1,..., \theta_b) \in \Theta}
\left| \sum_{b=1}^{B} \sqrt{\rho_b}\h_b^{\herm} \f_b e^{i\theta_b} \right|^2.
\end{align}
 After using the complementary phase sequences, the above SNR simplifies to
\begin{align}\label{eq:snr_jo_com}
\snr^{\joint-\comb} = \frac{B}{N_0} \sum_{b=1}^{B} \left| \sqrt{\rho_b}\h_b^{\herm} \f_b \right|^2.
\end{align}

The use of complementary phase patterns ensures that cross terms vanish during coherent combining, leaving only the sum of power terms from each BS.

For the fixed beamforming vectors at the BSs, let us evaluate the SNR gain of JT with complementary phases compared to the independent transmission of SSBs, as the latter is the approach commonly adopted in the current 5G NR standard. To ensure a fair comparison, we allocate the same number of transmission resources to both schemes. Specifically, we repeat the independent transmission $B$ times, matching the number of repetitions used in JT for the fixed beamformer used at the BSs. Assuming the UE is connected to its closest BS, denoted as $k$, the SNR with independent transmission repeated $B$ times is given by 
\begin{align} \label{eq:SNR_Independent}
\snr^{\independent} &= B \frac{\rho_k}{N_0} \left| \h_k^{\herm} \f_k \right|^2.
\end{align}
Considering that the same beam is used from the closest BSs for both joint and independent transmission, the SNR difference between JT and independent transmission, based on~\eqref{eq:snr_jo_com} and~\eqref{eq:SNR_Independent}, is given by
\begin{align} \label{eq:delta_SNR}
\Delta \snr &=10 \log_{10} \left( 1 + 
\frac{\sum_{ b \neq k} \left| \sqrt{\rho_b} \h_b^{\herm} \f_b \right|^2}{\left| \sqrt{\rho_k} \h_k^{\herm} \f_k \right|^2} \right) \ge 0.
\end{align}

It is important to note that the above analysis assumes the same number of resources is allocated for a given beamforming pattern at the BSs. However, in practice, the number of beamforming patterns required for JT may be significantly larger than that for independent transmission. Let $N^{\independent}$ denote the number of beamforming patterns used in independent SSB transmission at any BS. When considering all possible combinations across the $B$ cooperating BSs in JT, the total number of beamforming patterns becomes $N^{\joint} \triangleq (N^{\independent})^B$, which is generally much larger than in the independent case. To ensure a fair comparison in terms of resource usage including the beamforming patterns, the independent transmission can be repeated $R \triangleq \frac{B N^{\joint}}{N^{\independent}}$ times. With these additional repetitions, the SNR for independent SSB transmission in~\eqref{eq:SNR_Independent} can be rewritten as $\snr^{\independent} = R \frac{\rho_k}{N_0} \left| \h_k^{\herm}\f_k \right|^2$. Accordingly, the SNR gain defined in~\eqref{eq:delta_SNR} should be updated to reflect this as
\begin{align}\label{eq:delta_SNR2}
\Delta \snr \! = \! 10 \log_{10} \! \left( 1 \! + 
\frac{\sum_{ b \neq k} \left| \sqrt{\rho_b}\h_b^{\herm} \f_b \right|^2}{\left| \sqrt{\rho_k}\h_k^{\herm} \f_k \right|^2} \right) \! + \! 10 \log_{10} \! \left( \frac{N^\independent}{N^\joint} \right).
\end{align}
From the above, we observe that if $N^{\joint} > N^{\independent}$, the SNR gain from JT diminishes. In fact, it is even possible for independent SSB transmission to outperform JT under such conditions. In the following, we discuss the optimization of the number of beamforming combinations used in JT of SSB, i.e., $N^{\joint}$, to improve the SNR gain with the joint SSB transmission.

\subsection{Beam Selection}

In JT of SSBs, each location ideally benefits from a joint beam that maximizes the received SNR at that specific point. While using all $(N^{\independent})^B$ joint beam combinations would maximize the first term in~\eqref{eq:delta_SNR2}, it would also result in a prohibitively large number of SSB transmissions and significantly increase signaling overhead. Conversely, selecting a smaller number of joint beams reduces the total number of transmissions (i.e., the second term in~\eqref{eq:delta_SNR2}), but may lead to a suboptimal power gain due to less favorable beamforming configurations. To balance this trade-off, we select $N^{\joint} = N^{\independent}$ carefully chosen joint beam combinations from the total $(N^{\independent})^B$ possibilities.    It should also be noted that each selected beamforming vector in the JT scheme is repeated $B$ times with different phase settings. Therefore, the total number of SSB transmissions becomes $B N^{\independent}$ in the joint SSB design case. This ensures that the SNR gain in~\eqref{eq:delta_SNR2} remains positive, demonstrating that the proposed JT of SSBs method consistently outperforms independent transmission, even under equal resource constraints. To this end, the optimization problem in~\eqref{eq:Opt} with the SNR of coherent combining of the fixed phases at the UE given in~\eqref{eq:snr_jo_com}  simplifies to 
\begin{equation}
\left\{
\begin{aligned}
    \max \  & \frac{1}{|\setG|} \sum_{g \in \setG} 
    u\left( \snr^{\joint-\comb}_{\max}
    - {\snr^{\refe}} \right), \\
    \text{s.t. } & \mathcal{J} \triangleq \left\{ (\f_1, \dots, \f_B) \mid \f_b \in \mathcal{F}_b \right\},
    \ |\mathcal{J}|= N^{\independent}.
\end{aligned}
\right.
\end{equation}
where $\snr^{\joint{-}\comb}_{\max} \triangleq \max_{\f_b} \snr^{\joint{-}\comb}$. To solve this optimization problem, we consider the total of $(N^{\independent})^B$ possible joint beamforming combinations, where each combination corresponds to one element of $(\mathbf{f}_1, \dots, \mathbf{f}_B) \in \mathcal{F}_1 \times \cdots \times \mathcal{F}_B$. From the  $(N^{\independent})^B$ beams, we select a set of $N^{\independent}$ joint beams using a greedy beam selection strategy in which, at each iteration, the beam combination that provides the maximum coverage is selected from the uncovered area. Specifically, at iteration $n$, the best beam is selected as follows
\begin{align}\label{eq:beam_sel}
   [\f^n_1,\dots, \f^n_B ]  = \! \argmax_{(\mathbf{f}_1, \dots, \mathbf{f}_B) \in \{\mathcal{F} \setminus  \mathcal{ \bar F} \}}   & \sum_{g \in \bar \setG}   u\left( \snr^{\joint-\comb}_{\max} \! \!- {\snr^{\refe}} \right),
\end{align}
where $ \mathcal{F} \triangleq \mathcal{F}_1 \times \cdots \times \mathcal{F}_B$. $\mathcal{ \bar F}$ denotes the set of previously selected beam combinations, and $\bar{\mathcal{G}}$ represents the remaining area not yet covered by them. This process continues until $N^{\independent}$ beam combinations have been selected.

\subsection{Enhanced Phase and Beam Selection} \label{subsec:Enhanced_Phase}

The JT from all the $B$ BS with complementary phase alignment guarantees constructive signal reception throughout the coverage area. However, this approach implicitly assumes that all $B$ BSs contribute significantly to the received signal at every UE location. In practice, this assumption rarely holds due to the distance-dependent path loss. Most of the SNR gain typically comes from the BSs nearest to the UE. For instance, as illustrated in Fig.~\ref{fig:NetworkTopology}, at point $A$, the four closest BSs contribute the majority of the received SNR. In contrast, at point $B$, $BS_2$ and $BS_4$ provide the dominant contribution, while the others have minimal impact. This suggests that the set of coordinating BSs can vary depending on the UE location. Motivated by this observation, we propose a more efficient strategy in which the number of coordinated BSs and phase repetitions is selectively reduced based on the relative contribution of each BS to the UE received signal. BSs with negligible channel gain are excluded from the phase design, as their influence on the combined SNR whether constructive or destructive, is minimal. To this end, we define the set of dominant BSs for a given joint beam $i$ as
\begin{align}
    \mathcal{D}_{i} \!= \!\left\{ \! b \in \{1, \dots, B\} \ \middle|  \left| \sqrt{\rho_b} \mathbf{h}_b^{\herm} \mathbf{f}^{i}_b \right|^2 \! \geq \alpha \max_j \left| \sqrt{\rho_j} \mathbf{h}_j^{\herm} \mathbf{f}^{i}_j \right|^2 \right\},
\end{align}
where $\alpha \in (0, 1]$ is a threshold parameter that determines which BSs significantly contribute to the received power of the UE. This condition ensures that only BSs with a sufficiently strong contribution are included in the phase alignment design. Accordingly, the number of complementary phase sequences can be reduced to the number of dominant BSs transmitting to that particular location, i.e., $|\Theta| = |\mathcal{D}_i|$.

Although the absolute SNR decreases with this strategy, the SNR reduction is primarily due to using fewer SSB repetitions. In the fixed BS design, each joint beam $(\mathbf{f}_1, \dots, \mathbf{f}_B)$ is transmitted $B$ times using all orthogonal phase patterns. In contrast, in this approach, some beam repeats fewer than $B$ times, resulting in a lower repetition gain. However, despite the drop in absolute SNR, the relative SNR remains nearly unchanged. This is because, to ensure a fair comparison, the number of repetitions in the independent case is also reduced proportionally to match the total SSB budget, which similarly lowers its absolute SNR. As a result, the ratio between the two remains stable. Minor fluctuations may still occur due to the exclusion of weaker BSs in the phase design, whose constructive or destructive contributions are ignored. Nonetheless, these effects are expected to be small, as the omitted BSs have negligible power contributions.

\section{Simulation Results}
To evaluate the performance of the proposed scheme, we consider an operating carrier frequency of $7.5$ GHz. The setup includes $B = 4$ BSs positioned at the corners of a $100 \times 100~\text{m}^2$ square grid. Each BS is equipped with $4$ transmit antennas and uses a DFT-based beamforming codebook. While this setup is chosen for demonstration clarity, the approach is fully scalable; the number of antennas per BS can be increased without loss of generality, and the proposed beam and phase selection strategy and all the results remain valid across a wide range of practical configurations. Moreover, for visualization purposes, the simulation grid is scaled by a factor of $100$ relative to the wavelength, allowing clearer spatial interpretation of the results. Lastly, each BS transmits with a power of $\{\rho_b\}= 0$ dBm, and AWGN at the UE, $ N_0=-95$ dBm. 
 
 

We assume that in the independent scenario, each BS transmits $N^{\independent} = 4$ beams, selected from a DFT codebook of size $4 \times 4$. We consider the same number of beams for the JT case, where the selected beams follow the greedy approach described in Section III.B and are given by
\begin{equation}
\mathcal{J} =
\begin{bmatrix}
1 & 1 & 1 & 1 \\
2 & 4 & 2 & 4 \\
3 & 3 & 3 & 3 \\
4 & 2 & 4 & 2 
\end{bmatrix}, \label{eq:SelectedBeams}
\end{equation}
where each row corresponds to a beam index from the DFT codebook and each column represents a BS index. In JT, each selected beam is repeated $4$ times with different phase values, according to the phase set defined in~\eqref{eq:4Tuple}. Consequently, the total number of SSB transmissions in the JT case is $16$, accounting for all combinations of beams and phases. To ensure a fair comparison, we also assume that in the independent case, each of the $4$ beams is transmitted $4$ times.

\begin{figure*}[!t]
    \centering
    \begin{minipage}{0.65\linewidth}
        \centering
        \begin{minipage}{0.48\linewidth}
            \centering
            \includegraphics[width=0.9\linewidth]{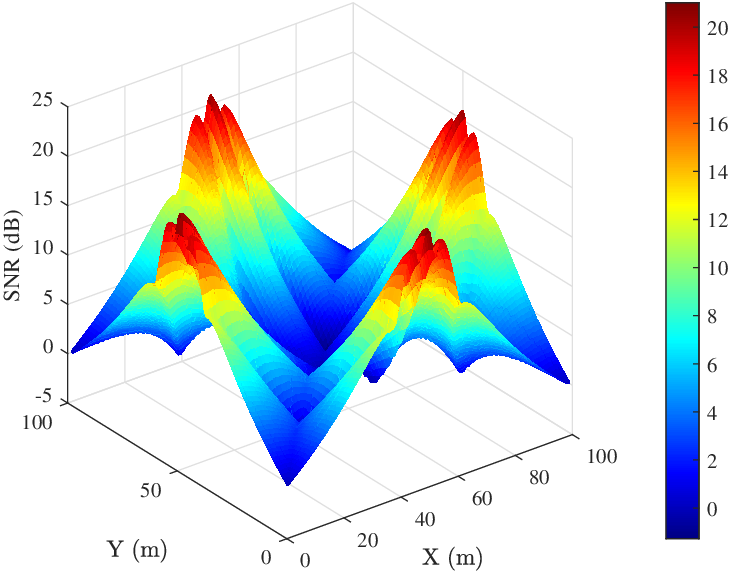}
            \caption*{\small (a) Independent transmission.}
        \end{minipage}
        \hfill
        \begin{minipage}{0.48\linewidth}
            \centering
            \includegraphics[width=0.9\linewidth]{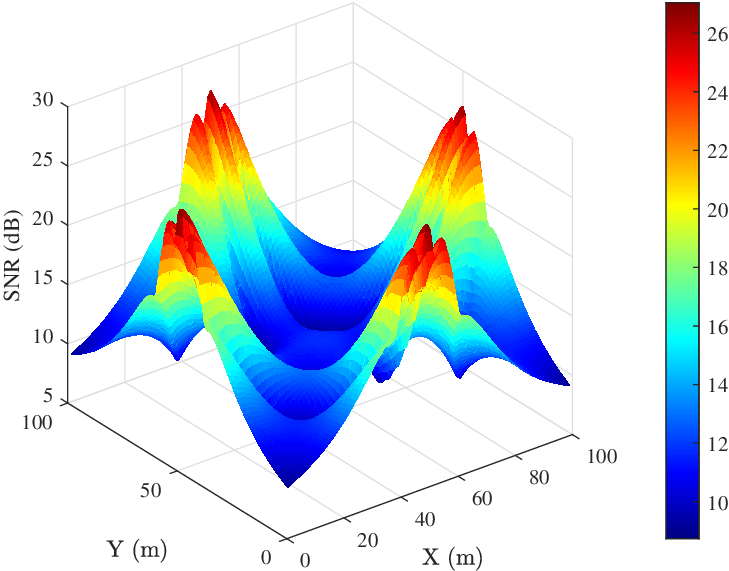}
            \caption*{\small (b) JT of SSBs with complementary phase sequences.}
        \end{minipage}
        \vspace{2mm}
        \caption{Absolute received SNR across the coverage area.}
        \label{fig:SNR_JT_Indep}
    \end{minipage}
    \hfill
    \begin{minipage}{0.32\linewidth}
        \centering
        \includegraphics[width=0.95\linewidth]{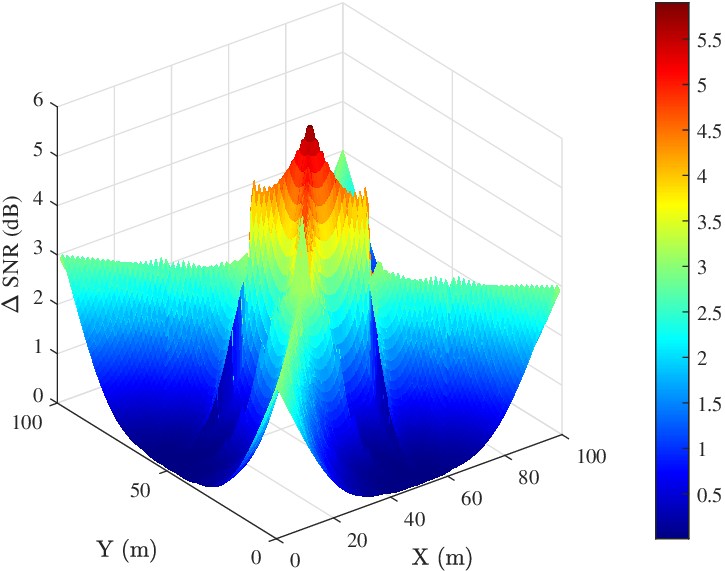}
        \vspace{2mm}
        \captionof{figure}{\small $\Delta\snr$ with the proposed JT of SSBs over independent transmission.}
        \label{fig:Delta4BS}
    \end{minipage}
    \vspace{-1mm}
\end{figure*}

The absolute $\snr$ values for different UE locations are shown in Fig.~\ref{fig:SNR_JT_Indep} for both independent and JT of SSBs. It is evident that JT consistently outperforms independent transmission. The absolute SNR is highest near the BSs and gradually decreases with increasing path loss. As expected, locations farther from all BSs exhibit the lowest SNR values due to high signal attenuation. Further insight can be drawn from Fig.~\ref{fig:Delta4BS}, which illustrates the SNR difference between joint and independent transmission, as defined in~\eqref{eq:delta_SNR2}. The performance gain from JT is most pronounced in the central region of the coverage area, where all BSs contribute more or less equally to the received signal. In contrast, at locations close to a single BS, the improvement is marginal, since the contribution from distant BSs is minimal due to higher path loss. Nonetheless, the absolute SNR at these locations is still relatively high compared to regions located far from any BS.

\begin{figure}[!b]
    \centering
    \includegraphics[width=0.8\linewidth, trim={0cm 0cm 0cm 0cm}, clip]{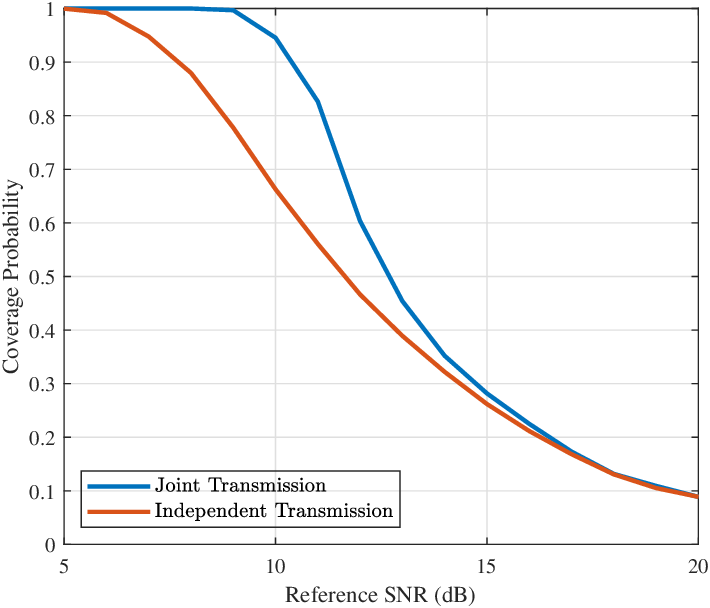}
    \caption{\small{Coverage percentage versus reference SNR.}}
    \label{fig:CoverageVsThreshold}
    \vspace{3mm}  
\end{figure}

To quantify the improvement in network coverage with JT of SSBs, we plotted the $\mathbb{P}(\snr^{\joint-\comb}_{\max} \ge \snr^{\refe} )$ in Fig.\ref{fig:CoverageVsThreshold} for different values of $\snr^{\refe}$. As shown, when the reference SNR threshold is very low (e.g., below $5$ dB), both joint and independent transmission schemes achieve $100\%$ coverage across the area. Conversely, when the threshold is too high (e.g., above $17$ dB), both methods cover only $16\%$ of the total area. The performance gap is most pronounced within the practical operating range, approximately between $5$ dB and $17$ dB. For instance, at $\snr^{\refe} = 10$ dB, the independent scheme provides coverage to approximately $66\%$ of the area, while the proposed JT-based design improves this to $94\%$. This gain highlights the effectiveness of the JT strategy in enhancing initial access performance.

\vspace{6mm}  
\captionsetup{skip=2mm} 

\begin{figure}[!b]
    \centering
    \includegraphics[width=0.85\linewidth]{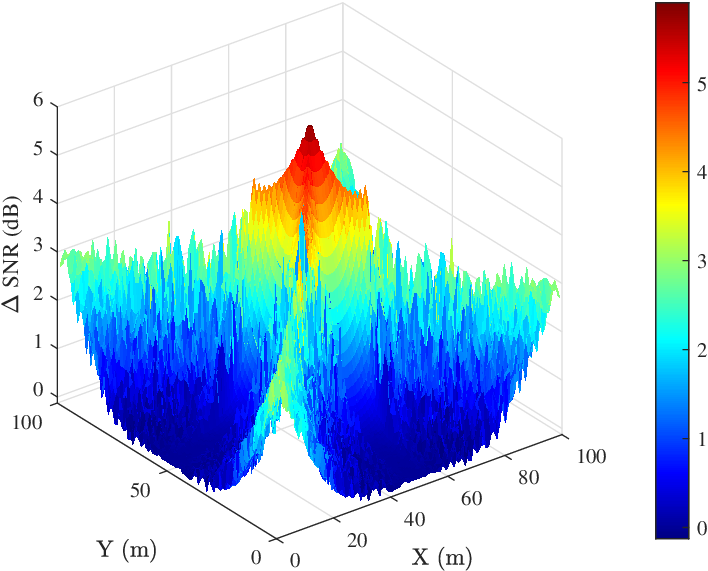}
    \caption{\small{$\Delta\snr$ with the enhanced phase and beam selection method.}}
    \label{fig:AsymGain}
    \vspace{10mm}  
\end{figure}

We now evaluate the performance of the enhanced location-based beam and phase selection strategy, which dynamically adjusts the number of cooperating BSs based on the UE position, as discussed in Section~\ref{subsec:Enhanced_Phase}. For example, in Fig.~\ref{fig:NetworkTopology}, at point $B$, only two BSs, $BS_2$ and $BS_3$, significantly contribute to the received signal and are therefore considered for joint SSB transmission. The optimal beams for each active BS are selected according to~\eqref{eq:SelectedBeams}, while the complementary phase sequences are adapted to the number of contributing BSs at each location. Therefore, at point $B$, since only two BSs are involved, the corresponding phase set is given as~\eqref{eq:2Tuple}, resulting in only two SSB repetitions for that region. This selective cooperation reduces the total number of SSB transmissions from 16 (with fixed BS cooperation) to 10. To ensure fairness, the independent transmission scheme is also adjusted to match the reduced transmission count; for instance, at point $B$, the SSB beam in the independent case is repeated only twice. This maintains equal resource usage between the two schemes. As illustrated in Fig.~\ref{fig:AsymGain}, the relative SNR gain remains nearly unchanged compared to the fixed number of BS cooperation scheme, despite the reduction in the number of transmissions. Minor fluctuations may occur due to the exclusion of weakly contributing BSs, but these have a negligible impact on overall performance due to their limited influence on the received signal power.

\vspace{6mm}  
\captionsetup{skip=2mm} 

\section{Conclusions}
This paper explored the use of JT for SSB transmission to enhance initial access coverage. We demonstrated that even without CSI at the transmitter, it is possible to design fixed phase sets that enable constructive reception at any UE location. To ensure a fair comparison with conventional methods, the number of joint beam combinations was set equal to the number of independent beams. Under this constraint, the proposed JT of the SSB method consistently outperforms independent SSB transmission. Additionally, by adapting the beam and phase configuration based on network geometry and the relative contributions of each BS, similar SNR gains can be achieved using fewer SSB transmissions. Overall, the proposed strategy provides a practical and resource-efficient approach to improving initial access coverage and can complement existing solutions in future network deployments.

\bibliographystyle{IEEEtran}
\bibliography{refs_abbr,refs}
\end{document}